\documentclass[twocolumn,showpacs,preprintnumbers,amssymb,aps]{revtex4}

\begin{document}

\title{Quasi-Local Energy Flux of Spacetime Perturbation}
\author{Roh-Suan Tung}\email{tung@shnu.edu.cn}
  \affiliation{Center for Astrophysics, Shanghai Normal University,
 Shanghai 200234, China}
\author{Hoi-Lai Yu}\email{hlyu@phys.sinica.edu.tw}
 \affiliation{Institute of Physics, Academia Sinica, Taipei 115, Taiwan ROC}

\date{October 14, 2008}

\begin{abstract}
A general expression for quasi-local energy flux for spacetime
perturbation is derived from covariant Hamiltonian formulation
using functional differentiability and symplectic structure
invariance, which is independent of the choice of the canonical
variables and the possible boundary terms one initially puts into
the Lagrangian in the diffeomorphism invariant theories. The
energy flux expression depends on a displacement vector field and
the 2-surface under consideration. We apply and test the
expression in Vaidya spacetime. At null infinity the expression
leads to the Bondi type energy flux obtained by Lindquist,
Schwartz and Misner. On dynamical horizons with a particular
choice of the displacement vector, it gives the area balance law
obtained by Ashtekar and Krishnan.
\end{abstract}

\pacs{04.20.Fy,04.70.Bw}%

\keywords{}

\maketitle

\section{Introduction}

It is well known that in General Relativity, because of
equivalence principle, locally we can not detect gravity and
therefore renders any unambiguous notion of local density for
conserved quantities impossible. On the other hand, the notion of
total conserved quantities for spatial and null infinity are well
understood. Therefore, there has been hope that one should be able
to find an appropriate notion of quasi-local conserved quantities
for finite spacetime domains \cite{Quasi}.

A systematic way to study conserved quantities is through
Hamiltonian. In a series of papers \cite{QL1, QL2, Nester, Tung2},
a covariant Hamiltonian formalism was developed to obtain
quasi-local energy-momentum expressions and found that the
Hamiltonian boundary term also determines the boundary conditions
for stationary spacetime.

In general, for a dynamical gravitating system, the gravitational
energy-momentum is not conserved, but in such case we would expect
a meaningful total or quasi-local energy flux expression. In
\cite{Wald2000}, covariant Noether charge (Hamiltonian)
formulation was first used to identify the Bondi energy flux. For
trapping  and dynamical horizons \cite{Hayward1994}, an energy
flux expression was also obtained by Ashtekar and Krishnan
\cite{Ashtekar-Krishnan2002}. The formalism were then applied to
trace the development of trapping horizons in black hole formation
and evaporation \cite{Hayward2006}. For general spacetime regions,
definitions of quasi-local energy flux was also developed using
covariant Hamiltonian formalism \cite{Nester}. The Bondi energy
flux was derived but the energy flux at dynamical horizons was not
addressed there. Much discussions using marginally trapped
surfaces and dynamical horizons were given in terms of the Vaidya
spacetimes \cite{Sch,Ben-Dov,B-S}.

In this paper, we shall construct explicitly a physically well
defined energy flux formula for general dynamical gravitating
system using the functional differentiability of the Hamiltonian
and symplectic structure invariance of the theory which is
independent of the choice of canonical variables and the boundary
terms that one initially puts into the Lagrangian. We use
spherical symmetric Vaidya spacetime as an example to test our
energy flux expression \cite{LSM}. The Bondi type energy flux at
the null infinity and the energy flux at the dynamical horizons
are also derived respectively in Schwarzschild and
Painlev\'{e}-Gullstrand coordinates.

In section II, we derive our energy flux expression within
covariant Hamiltonian formulation using functional
differentiability and symplectic structure invariance.  The
important features of this expression are the independence of the
choice of the canonical variables and the possible boundary terms
that one initially puts into the Lagrangian for diffeomorphism
invariant theories. In section III, we apply our expression to
General Relativity. Examples for Vaidya spacetime to obtain the
Bondi type energy flux at null infinity are then given in the
section IV. In there, the area dependence and implications of the
first law is also observed. In section V, we calculate the energy
flux for dynamical horizons and compare with the results obtained
by Ashtekar and Krishnan \cite{Ashtekar-Krishnan2002} for a
particular choice of the displacement vector. We discuss the
results in section VI.

\section{Noether conserved quantities and energy flux}

In the following, we shall apply the Noether charge approach
\cite{Wald,Wald1990,Wald2000,W2, Nester1991, Tung1, Tung2} for
dynamical spacetime using self consistence of functionally
differentiable Hamiltonian and invariance of the corresponding
symplectic structure to define a physical energy flux. Starting
with a diffeomorphism invariant first order Lagrangian 4-form  $
{\cal L} = {\cal L} (\phi,p) = d \varphi \wedge p -
\Lambda(\varphi,p)$, where $\phi$ denotes an arbitrary collection
of dynamical fields, and $p$ being the corresponding conjugate
momentum \cite{Nester1991}. The field equations, $\frac{\delta\cal
L}{\delta\phi}=0$, $\frac{\delta\cal L}{\delta p}=0$ are obtained
by computing the first variation of the Lagrangian,
\begin{equation} \label{4}
\delta {\cal L}= d (\delta\phi\wedge p ) + \delta \phi \wedge
{\delta {\cal L}\over \delta \phi}  +  {\delta {\cal L}\over \delta
p} \wedge \delta p,
\end{equation}
where $\delta\phi\wedge p$ is the symplectic potential 3-form. For
any diffeomorphism generated by a smooth vector field $\xi$, we can
replace the variational derivative $\delta$ by the Lie derivative
$\pounds_\xi$
\begin{equation} \label{5}
\pounds_\xi {\cal L} = d(\pounds_\xi \phi \wedge p) + \pounds_\xi
\phi \wedge {\delta {\cal L}\over \delta \phi}  +  {\delta {\cal
L}\over \delta p} \wedge \pounds_\xi p,
\end{equation}
Using the identity $\pounds_\xi = i_\xi d+ d i_\xi$  and replacing
$d i_\xi {\cal L}$  with $\pounds_\xi {\cal L}$, one can define a
conserved Noether current 3-form (or Hamiltonian 3-form
\cite{Nester}) ${\cal H}(\xi)$ by
\begin{equation}
{\cal H}(\xi)=\pounds_\xi \phi \wedge p -i_\xi {\cal L}(\phi) ,
\end{equation}
such that by equation (\ref{5}) the Noether current ${\cal H}(\xi)$
is closed ($d{\cal H}(\xi)=-\pounds_\xi \phi \wedge {\delta {\cal
L}\over \delta \phi}  - {\delta {\cal L}\over \delta p} \wedge
\pounds_\xi p\simeq 0$) when the field equations are satisfied.
Locally there exist a 2-form $Q(\xi)$ (called the Noether charge)
such that ${\cal H}(\xi)=d Q(\xi) +$ ``field equation terms'' or in
general,
\begin{equation}
{\cal H}(\xi) =: \xi^\mu {\cal H}_\mu + d {Q}(\xi).
\end{equation}
Note that here ${\cal H}_\mu$ are constraints including matter
fields contribution. When integrated on a 3-space $\Sigma$, it
gives a Hamiltonian
\begin{equation}\label{8}
H (\xi)=\int_\Sigma {\cal H} (\xi)= \int_\Sigma \xi^\mu {\cal
H}_\mu +\oint_{\partial\Sigma} Q,
\end{equation}
therefore, $Q$ can also be interpreted as the boundary term $B =
\oint_{\partial\Sigma} Q$ which defines the value of the
Hamiltonian. We want to stress here, although $H$ is called
Hamiltonian in the literature, it may not be functionally
differentiable to define conserved quantities along the
displacement vector field $\xi$ which generate diffeomorphism
invariant transformations. Following the work of Regge and
Teitelboim \cite{ReggeTeitelboim1974}, we justify the functional
differentiability of $H$ as the total Hamiltonian by further
varying $H$,
\begin{equation}
\delta H(\xi)=\int_{\Sigma} \delta {\cal H}(\xi)=\int_{\Sigma}
\delta(\pounds_\xi \phi \wedge p) -i_\xi \delta {\cal L}(\phi).
\end{equation}
Using equation (\ref{4}) and dropping the field equation terms, we
arrive at
\begin{eqnarray}
\delta H(\xi)&=& \int_{\Sigma} \delta(\pounds_\xi \phi \wedge p) -
\pounds_\xi (\delta\phi\wedge p )  + \oint_{\partial\Sigma} i_\xi
(\delta\phi\wedge p)  \nonumber\\ \label{10} &=& \int_{\Sigma}
(\pounds_\xi \phi \wedge \delta p - \delta\phi\wedge \pounds_\xi p
)  + \oint_{\partial\Sigma} i_\xi
(\delta\phi\wedge p), \nonumber\\
\end{eqnarray}
and the symplectic 3-form $\Omega$ is defined by
\begin{equation}
\Omega(\delta_1,\delta_2)=\int_{\Sigma} \delta_1 \phi \wedge
\delta_2 p- \delta_2 \phi\wedge \delta_1 p.
\end{equation}
There can be two possibilities. If $ \oint_{\partial\Sigma} i_\xi
(\delta\phi\wedge p)=0$ then $H(\xi)$ is automatically
functionally differentiable, conserved along the vector field
$\xi$, i.e. $\pounds_\xi {H}(\xi)=0$. When one can find boundary
conditions (i.e. see reference \cite{Nester}) to give $
\oint_{\partial\Sigma} i_\xi (\delta\phi\wedge p)=\delta
\widetilde{B}(\xi)$, to modify $H$ to $\widetilde{H} =
H(\xi)-\widetilde{B}(\xi)$ such that $\widetilde{H}$ is still
functionally differentiable, conserved along the vector field
$\xi$, i.e. $\pounds_\xi \widetilde{H}(\xi)=0$. The function
${H}(\xi)$ and $\widetilde{H}(\xi)$ are called the functionally
differentiable Hamiltonian conjugate to $\xi$
\cite{Wald1990,Nester1991,Wald,Wald2000,Nester,Tung2}.

In general dynamical gravitating systems, when spacetime is
non-stationary, there does not exist boundary conditions to
achieve a functionally differentiable Hamiltonian to define
conserved quantities. However, the replacement  of $\delta$ by
$\pounds_\xi$ in equation(\ref{10}) will still lead to the
following flux expression,
\begin{equation}\label{12}
\pounds_\xi H(\xi) =  \oint_{\partial\Sigma} i_\xi
(\pounds_\xi\phi\wedge p).
\end{equation}
But this flux expression has the ambiguity on the choice of the
canonical variables between $\phi$ and $p$. Following
\cite{Wald2000}, a prescription was developed by insisting on the
self consistency of functionally differentiable Hamiltonian and
invariance of symplectic structure(symmetric under $-\phi$ and $p$
interchanges) \cite{Tung2001}.

To proceed, we perform a second variation $\Delta$ on $\pounds_\xi
H$ in equation (\ref{12}), and identify the energy flux as in the
followings. We have,
\begin{equation}
\Delta \pounds_\xi H (\xi) = \oint_{\partial\Sigma}
i_\xi(\pounds_\xi \Delta  \phi \wedge p +\pounds_\xi \phi \wedge
\Delta p ),
\end{equation}
where $\Delta \pounds_\xi=\pounds_\xi \Delta$ is assumed. We
subtract a Lie derivative term on both sides,
\begin{equation} \label{14a}
\Delta \pounds_\xi H(\xi) -\pounds_\xi \oint_{\partial\Sigma}
i_\xi (\Delta  \phi \wedge p) =  \oint_{\partial\Sigma} i_\xi
(\pounds_\xi  \phi \wedge \Delta p -\Delta \phi \wedge \pounds_\xi
p),
\end{equation}
and observe that the last term has a symplectic structure and
therefore allows us to define a quantity $\mathbb{E}(\xi)$ by
\begin{equation}
\mathbb{E}(\xi)= \Delta H(\xi)- \oint_{\partial\Sigma} i_\xi
(\Delta \phi \wedge p).
\end{equation}
Note that the above equation (\ref{14a}) bears the same form as
equation (\ref{10}), we arrive at the first main result of this work,
\begin{equation}\label{15}
\mathbb{F}(\xi)=\pounds_\xi \mathbb{E}(\xi)=
\oint_{\partial\Sigma} i_\xi (\pounds_\xi \phi \wedge \Delta p
-\Delta  \phi \wedge \pounds_\xi p).
\end{equation} 
From the surface integral form of $\mathbb{F}(\xi)$,
non-conserving nature and only being functional differentiable to
generate the correct dynamical evolution on the surface
$S=\partial\Sigma$, we can therefore interpret it being the total
energy flux across some dynamical surface boundary $S$. To further
confirm this energy flux interpretation, we observe another
important feature of this flux expression, namely, the symplectic
structure invariant character will automatically gives the correct
boundary terms, boundary conditions and a conserved Hamiltonian as
discussed in \cite{Nester} in the stationary cases.

Note that our general expression $\mathbb{F}(\xi)$ is precisely the sum of the
two flux expressions of [4],
\begin{eqnarray}
\mathbb{F}(\xi)
&=&\mathbb{F}_{\rm Dirichlet}(\xi) + \mathbb{F}_{\rm Neumann}(\xi)  \nonumber\\
&=&\mathbb{F}_{\rm dynamic}(\xi) + \mathbb{F}_{\rm constraint}(\xi) ,
\end{eqnarray} 
where $\mathbb{F}_{\rm Dirichlet}(\xi)$, $\mathbb{F}_{\rm Neumann}(\xi)$,
$\mathbb{F}_{\rm dynamic}(\xi)$ and $\mathbb{F}_{\rm constraint}(\xi)$ are
the flux expressions with certain variables being fixed on the boundary
(see Appendix).
For a general dynamic spacetime, we want to allow our symplectic
variables to be completely dynamical without any variable being fixed
on the boundary. An example is
dynamical black holes which we shall discuss in section V.

We would also like to point out that it is the
functional differentiability and the  symplectic structure
invariance which  frees the energy flux expression from the
ambiguities in determining  the boundary terms one initially put
into the Lagrangian and therefore allows us to define the energy
flux expression uniquely.

\section{Energy flux from Spacetime Perturbations}

We now apply the prescription described in the previous section to
General Relativity,
\begin{equation}
 S=\int_{\Sigma} {\cal L}={1\over 16\pi} \int_{\Sigma} R^{ab}\wedge
\left(  \ast(\vartheta_a\wedge\vartheta_b) \right)+ {\cal
L}_{matter},
 \end{equation}
where $R^{ab}=d\omega^{ab}+\omega^a{}_c \wedge \omega^{cb}$ is the
curvature 2-form constructed from the connection 1-form $\omega^{ab}$,
 $\ast(\vartheta^a\wedge\vartheta^b)=\frac{1}{2}
 \epsilon^{ab}{}_{cd}\vartheta^c\wedge\vartheta^d$, and $g=
 \eta_{ab}\, \vartheta^a \otimes_s \vartheta^b$ is the metric,
 where $\eta_{ab}={\rm diag}(-1,1,1,1)$ and
 $\vartheta^a$ is the orthonormal
frame 1-form field. It is important to note that one can add a
boundary term to the above action, however, adding such a boundary
term will not change the result using our symplectic structure
invariant prescription described in the previous section.

Denote $\eta_{ab}=
 (\ast(\vartheta_a\wedge\vartheta_b))
 =\frac{1}{2}\epsilon_{abcd}\vartheta^c \wedge \vartheta^d $,
 $\omega^{ab}$
being the spin connection that solves the equation of motion.
The expression of Noether current 3-form in equation(\ref{8}) is given by
\begin{eqnarray}\label{20}
H(\xi) &=& \int_{\Sigma}{\cal H}(\xi)=\int_{\Sigma}  d Q(\xi) =
{1\over 16\pi}\oint_{\partial\Sigma}
i_\xi \omega^{ab}\wedge \eta_{ab}  , \nonumber\\
\end{eqnarray}
where $Q(\xi)$ is the Noether charge 2-form appears as a total
derivative therefore can also be interpreted as the boundary term,
here,  we assume the field equations are satisfied.  For
stationary spacetime but rather generic situation
($\oint_{\partial\Sigma} i_\xi \phi\wedge p\neq 0$), we require
\begin{eqnarray}
\delta \widetilde{H}&=&\delta (H-\widetilde{B}) \nonumber\\
&=& {1\over 16\pi} \left[\delta \oint_{\partial\Sigma} i_\xi
\omega^{ab}\wedge \eta_{ab}- \oint_{\partial\Sigma}
i_\xi (\delta\omega^{ab} \wedge \eta_{ab})\right] \nonumber\\
&=& {1\over 16\pi} \left[ \oint_{\partial\Sigma} i_\xi \omega^{ab}
\wedge \delta \eta_{ab} + \delta \omega^{ab}\wedge i_\xi
\eta_{ab}\right] =0.
\nonumber\\
\end{eqnarray}

This can be satisfied by the boundary conditions on a bifurcate
Killing horizons for stationary black holes \cite{W2, Wald,
Wald1990} such that $\pounds_\xi \vartheta^a = 0,  \pounds_\xi
\omega^{ab}=0$. In this case the energy flux is zero, $\pounds_\xi
\widetilde{H}=0$ and from equation (\ref{20}) we can obtain the
Noether charge for stationary case.

For dynamical cases, following equation (\ref{15}) we obtain the
corresponding energy flux formula using the previous described
invariant symplectic structure prescription,
\begin{equation}\label{Master}
\mathbb{F}(\xi)= {1\over 16\pi}\oint_{\partial\Sigma} i_\xi \left(
\pounds_\xi \omega^{ab} \wedge \Delta \eta_{ab} -\Delta
\omega^{ab} \wedge \pounds_\xi \eta_{ab} \right) .
\end{equation}
For small perturbation away from stationary spacetime, we can
define $\Delta\omega=\omega_{dynamical}-\omega_{stationary}$, and
$\Delta \eta=\eta_{dynamical}-\eta_{stationary}$. This is a
consistent definition, as when in the stationary limit, the
expression will give zero flux, correct boundary conditions and
correct value of the Noether charges for stationary black holes.

\section{Vaidya spacetime example, Bondi type energy flux and First law}

As an example for calculations of our energy flux expression,
we consider the Vaidya spacetime which describes a spherically
symmetric collapse of null dust (radiation). The metric is given by
\begin{equation}
ds^2=- e^{2\psi} dt^2 +e^{-2\psi} d r^2 + r^2(d\theta^2+\sin^2\theta
d\phi^2),
\end{equation}
where $\psi=\psi(t,r)$ and $e^{2\psi}=1-2m(t,r)/ r$.

In this coordinate, the marginally
trapped surfaces are given by $r=2 m(t,r)$. For constant $m(t,r)$,
this is just the standard Schwarzschild metric. Now consider a
perturbation $\Delta m(t,r)$ away from the stationary solution,
\begin{equation}
m(t,r)=m_{0}+\Delta m (t,r),
\end{equation}
because $m_0$ is a constant, this implies $m'=\partial_r (\Delta m)$
and $\dot{m}=\partial_t (\Delta m)$. In terms of the orthonormal
frames, the natural choice is,
\begin{eqnarray}
\vartheta^0=e^\psi dt, &\quad& \vartheta^1= e^{-\psi} dr,
\nonumber\\
\vartheta^2=rd\theta, &\quad& \vartheta^3=r\sin\theta d\phi ,
\end{eqnarray}
with corresponding basis vectors:
\begin{eqnarray}
e_0= e^{-\psi} \partial_t, &\quad& e_1= e^\psi \partial_r, \nonumber\\
e_2={1\over r} \partial_\theta, &\quad& e_3={1\over r\sin\theta}
\partial_\phi.
\end{eqnarray}

For $\xi=c_1\partial_t+c_2 \partial_r$, the Lie
derivatives of $\vartheta^a$ are
\begin{eqnarray}
\pounds_\xi \vartheta^0&=& {-1\over r e^{\psi}}
\left[c_1 \dot{m} +c_2 \left(m'- {m\over r}\right)\right] dt,\nonumber\\
\pounds_\xi \vartheta^1&=& {1\over  r e^{3\psi}}
\left[c_1 \dot{m} +c_2 \left(m'- {m\over r}\right)\right]  dr,\nonumber\\
\pounds_\xi \vartheta^2 &=& c_2 d\theta, \nonumber\\
\pounds_\xi \vartheta^3 &=& c_2 \sin\theta d\phi.
\end{eqnarray}
The spin-connection $\omega^{ab}$ has the following nonvanishing
terms:
\begin{eqnarray}
\omega^{01}&=& {1 \over r e^{4\psi}} \dot{m} dr
- {1\over r}\left(m'-{m\over r}\right) dt
=-\omega^{10}, \nonumber\\
\omega^{12}&=& - e^\psi d\theta =-\omega^{21}, \nonumber\\
\omega^{13}&=&- e^\psi \sin\theta d\phi =-\omega^{31},  \nonumber\\
\omega^{23}&=& - \cos\theta d\phi =-\omega^{32}.
\end{eqnarray}
The corresponding Lie derivatives of $\omega^{ab}$ have the follow
nonvanishing terms,
\begin{eqnarray}
\pounds_\xi\omega^{01} &=& {1 \over e^{4\psi}} \left[ {c_1\over r}
\left( \ddot{m}
+{4\dot{m}^2\over r e^{2\psi}} \right) \right] dr \nonumber\\
 && +{1 \over e^{4\psi}} \left[ {c_2\over r}
\left(\dot{m}'+{4 \dot{m} m' \over r e^{2\psi}}-{\dot{m}\over r}-{4
m \dot{m} \over r^2 e^{2\psi}}\right)
 \right]
 dr
\nonumber\\
&& - \left[{c_1\over r} \left( \dot{m}'- {\dot{m}\over r} \right)
\right] dt \nonumber\\&& - \left[{c_2\over r}\left(m''-{2m'\over
r}+{2m\over
r^2}\right)\right] dt \nonumber\\&=&-\pounds_\xi\omega^{10}, \nonumber\\
\pounds_\xi\omega^{12}&=& {1\over r e^{\psi}} \left[c_1 \dot{m} +c_2
\left(m'- {m\over r}\right)\right] d\theta
 \nonumber\\
 &=&-\pounds_\xi\omega^{21},\nonumber\\
 \pounds_\xi\omega^{13}&=& {1\over r e^{\psi}} \left[c_1 \dot{m}
 +c_2 \left(m'- {m\over r}\right)\right] \sin\theta d\phi \nonumber\\
 &=& -\pounds_\xi\omega^{31},  \nonumber\\
 \pounds_\xi\omega^{23}&=& 0=-\pounds_\xi\omega^{32}.
\end{eqnarray}
The perturbation of the orthonormal tetrad the spin-connection have
the following forms respectively,
\begin{eqnarray}
\Delta \vartheta^0&=& -{1 \over r e^{\psi}} {\Delta m} \, dt, \nonumber\\
\Delta \vartheta^1&=& {1\over r e^{3\psi}}  {\Delta m} \, dr, \nonumber\\
\Delta \vartheta^2 &=& 0, \nonumber\\
\Delta \vartheta^3 &=& 0,
\end{eqnarray}
\begin{eqnarray}
\Delta\omega^{01}&=& \left({4 \dot{m} {\Delta m} \over r e^{3\psi}
}  + {1\over r e^{4\psi}} \Delta \dot{m} \right) dr \nonumber\\
&& - {1\over r} \left( \Delta m'- {\Delta m\over r} \right) dt
 \nonumber\\
 &=&-\Delta\omega^{10} , \nonumber\\
\Delta\omega^{12}&=&{1\over r e^{\psi}} \Delta m \, d\theta
=-\Delta\omega^{21} ,\nonumber\\
\Delta\omega^{13}&=&{1\over r e^{\psi}} \Delta m \, \sin\theta d\phi
=-\Delta\omega^{31} ,\nonumber\\
\Delta\omega^{23}&=&0 =-\Delta\omega^{32} .
\end{eqnarray}

Many of the terms vanish in the energy flux $\mathbb{F}(\xi)$ ,
the remaining nonvanishing terms that will contribute are,
\begin{eqnarray}
&&{1\over 8\pi} \oint_{\partial\Sigma} i_\xi
(\pounds_\xi\omega^{12}\wedge \Delta
\eta_{12}+\pounds_\xi\omega^{13}\wedge \Delta \eta_{13})
\nonumber\\
&=&- c_1 {\Delta m \over re^{2\psi}} \left[ c_1 \dot{m} +c_2
\left(m'- {m\over r}\right)\right] ,
\end{eqnarray}
\begin{eqnarray}
&&-{1\over 8\pi} \oint_{\partial\Sigma} i_\xi (\Delta
\omega^{12}\wedge \pounds_\xi \eta_{12} +\Delta
\omega^{13}\wedge \pounds_\xi \eta_{13}) \nonumber\\
&=& c_1 \left[ {\Delta m \over re^{2\psi}} \left[ c_1 \dot{m} +c_2
\left(m'- {m\over r}\right)\right]- c_2 {\Delta m\over r} \right],
\nonumber\\
\end{eqnarray}
and
\begin{eqnarray}
&&- {1\over 8\pi} \oint_{\partial\Sigma} i_\xi
(\Delta\omega^{01}\wedge \pounds_\xi \eta_{01}) \nonumber\\
 &=& c_2 \left[ \left( \Delta m'- {\Delta m\over r} \right)
c_1 -\left({4 \dot{m} {\Delta m} \over e^{3\psi} } + {\Delta
\dot{m}\over e^{4\psi}} \right) c_2
 \right]. \nonumber\\
\end{eqnarray}

Finally, the total energy flux is
\begin{eqnarray}\label{27}
\mathbb{F}(\xi) &=& c_2 \left[ \left( \Delta m'- {2 \Delta m\over
r} \right) c_1 -\left({4 \dot{m} {\Delta m} \over e^{3\psi} } +
{\Delta \dot{m}\over e^{4\psi}} \right) c_2
 \right].\nonumber\\
\end{eqnarray}

Taking $u=t-r=const$ and $t, r \rightarrow \infty$ to approach the
 null infinity and dropping the term contains $\dot{m} \Delta m$
 which is of higher order in $\Delta$, we arrive at the
Bondi type energy flux
\begin{equation} \label{bondi}
\mathbb{F}(\xi) = c_2 (c_1 m' - c_2 \dot{m})= -
\partial_u m(u),
\end{equation}
where we have put $c_1=1$ and $c_2=1$ by requiring that $\xi$ and
$\Delta$ defines the same direction of mass changes for the
consistency of interchanging $\Delta$ and $\pounds_\xi$. The same
energy flux result, equation (\ref{bondi}), was also obtained long
ago by Lindquist, Schwartz and Misner \cite{LSM} using
Landau-Lifshitz stree-energy pseudotensor. Such energy flux
$-\partial_u m(u)$ has the interpretation as the luminosity of the
star as seen by an observer at null infinity.

It is interesting to notice that the nonvanishing term come only
from the following equations
\begin{eqnarray}
\mathbb{F}(\xi) &=&-{1\over 16\pi}\oint_{\partial\Sigma} i_\xi
[\Delta \omega^{01}] \wedge \pounds_\xi
(\vartheta^2\wedge \vartheta^3) \nonumber\\
&& -{1\over 16\pi}\oint_{\partial\Sigma} (i_\xi
\vartheta^1)[\Delta\omega^{12}\wedge\pounds_\xi\vartheta^3
 -\Delta\omega^{13}\wedge\pounds_\xi\vartheta^2]
\nonumber\\
&=&{1\over 16\pi}\oint_{\partial\Sigma} i_\xi [-\Delta \omega^{01}
+ {\Delta m \over r^2 e^\psi}\vartheta^1] \wedge \pounds_\xi
(\vartheta^2\wedge \vartheta^3) , \nonumber\\
\end{eqnarray}
where $\vartheta^2 \wedge \vartheta^3$ is the area element. This
indicates the first law for general spacetime regions. For
stationary spacetimes where $\pounds_\xi
(\vartheta^2\wedge\vartheta^3)=0$, the energy flux vanishes. The
appearance of the first law provides a nontrivial consistent check
of our energy flux expression.

\section{Energy flux from Dynamical Horizons: Painlev\'{e}-Gullstrand
coordinates }

On dynamical black-hole horizons, $r=2m$, which implies $m'={1/2},
\Delta m'={1/2}, m/r=1/2$. The first term of the above total energy
flux (\ref{27}) becomes
\begin{eqnarray}
\mathbb{F}(\xi)&=& \left[ {c_1 c_2\over 2}  \left( 1- {4 \Delta
m\over r} \right)\right]^{r_2}_{r_1}.
 \nonumber\\
\end{eqnarray}
Note that unlike in the Bondi type energy flux case which is defined at
the null infinity 2-sphere boundary, here for dynamical horizons,
 $\Sigma$ is bounded by two cross
sections, $\partial\Sigma=\partial\Sigma_1+\partial\Sigma_2$, with
radius changes from  $r_1$ to  $r_2$ dynamically because of the
outgoing energy flux. However, the second term in (\ref{27}) is
singular on dynamical horizons where $e^\psi=0$
(although $\dot{m}=0$) because of the coordinate singularity on the horizons.

In order to study the energy flux in dynamical horizons, we make a
coordinate transformation to the Painlev\'{e}-Gullstrand time
coordinate $T$ which is related to the Schwarzschild coordinate $t$
by,
\begin{equation}
T=t+4m\left[ \sqrt{r\over 2m} + {1\over2} \ln \left( {\sqrt{r\over
2m}-1 \over \sqrt{r\over 2m}+1} \right)\right].
\end{equation}
In terms of the Painlev\'{e}-Gullstrand coordinates, the metric can
be written as
\begin{equation}
ds^2=-dT^2 +(dr+ \sqrt{2m(T, r)\over r} d T)^2 +
r^2(d\theta^2+\sin^2\theta d\phi^2).
\end{equation}
with the choice of the orthonormal frames,
\begin{eqnarray}
\vartheta^0=dT, &\quad& \vartheta^1= dr +\sqrt{2m(T,r)\over r} dT,
\nonumber\\
\vartheta^2=rd\theta, &\quad& \vartheta^3=r\sin\theta d\phi ,
\end{eqnarray}
and the corresponding basis vectors,
\begin{eqnarray}
e_0= \partial_T - \sqrt{2m(T,r)\over r} \partial_r, &\quad&
e_1=\partial_r, \nonumber\\ e_2={1\over r} \partial_\theta,
\qquad\qquad\qquad &\quad& e_3={1\over r\sin\theta}
\partial_\phi.
\end{eqnarray}

In order to compare results for dynamical horizons obtained by
Ashtekar and Krishnan \cite{Ashtekar-Krishnan2002} previously, we
choose an $r$ dependent $\tilde {c}_2$ in the displacement vector,
$\xi=\tilde {c}_1 \partial_T +\tilde{c}_2(r)\partial_r$ instead of
a constant $\tilde {c}_2$.  We can now calculate the Lie
derivatives of $\vartheta^a$ and obtain
\begin{eqnarray}
\pounds_\xi \vartheta^0&=& 0, \nonumber\\
\pounds_\xi \vartheta^1&=&{1\over\sqrt{2mr}}\left[ \tilde {c}_1 \dot{m}
+\tilde{c}_2(r) \left( m'- {m\over r}\right)\right] dT + \tilde{c}_2'(r)dr,\nonumber\\
\pounds_\xi \vartheta^2 &=& \tilde{c}_2(r) d\theta, \nonumber\\
\pounds_\xi \vartheta^3 &=& \tilde{c}_2(r) \sin\theta d\phi.
\end{eqnarray}
The corresponding spin-connection $\omega^{ab}$ will have the following nonvanishing
terms:
\begin{eqnarray}
\omega^{01}&=&-  (m'- {m\over r}) \left({1\over \sqrt{2mr}} dr
+{1 \over r} dT \right)=-\omega^{10}, \nonumber\\
\omega^{02}&=&- {\sqrt{2m \over r}} d\theta = -\omega^{20}, \nonumber\\
\omega^{03}&=&- {\sqrt{2m \over r}} \sin\theta d\phi = -\omega^{30}, \nonumber\\
\omega^{12}&=& - d\theta =  -\omega^{21}, \nonumber\\
\omega^{13}&=& - \sin\theta d\phi =  -\omega^{31}, \nonumber\\
\omega^{23}&=& - \cos\theta d\phi =-\omega^{32},
\end{eqnarray}
and the Lie derivatives of $\omega^{ab}$ have the follow nonvanishing terms:
\begin{eqnarray}
\pounds_\xi\omega^{01}&=&
{\tilde {c}_1\over\sqrt{2mr}}(-\dot{m}'+{\dot{m}m'\over 2m} +{\dot{m}\over2r})dr
 \nonumber\\
&&
 +{\tilde{c}_2(r)\over\sqrt{2mr}}(-m''+{(m')^2\over 2m}+{m'\over r} -{3m\over 2
 r^2})dr \nonumber\\
&& + {\tilde{c}_2'(r)\over \sqrt{2mr}}({m\over r }-m')dr
+\tilde {c}_1({\dot{m}\over r^2} -{\dot{m}' \over r}) dT \nonumber\\
&& + \tilde{c}_2(r) (2{m'\over r^2}-2{m\over r^3}-{m'' \over r}) dT \nonumber\\
&=&-\pounds_\xi\omega^{10}, \nonumber\\
 \pounds_\xi\omega^{02}&=&
- {1\over \sqrt{2mr}} \left[ \tilde {c}_1 \dot{m} +\tilde{c}_2(r)
\left(m'-{m\over r}\right)\right]   d\theta
 \nonumber\\
  &=& -\pounds_\xi\omega^{20}, \nonumber\\
\pounds_\xi\omega^{03}&=&
 - {1\over \sqrt{2mr}} \left[ \tilde {c}_1 \dot{m}
 +\tilde{c}_2(r) \left(m'-{m\over r}\right)\right]
 \sin\theta d\phi \nonumber\\ &=&-\pounds_\xi\omega^{30}, \nonumber\\
\pounds_\xi\omega^{12}&=& 0 = -\pounds_\xi\omega^{21}, \nonumber\\
\pounds_\xi\omega^{13}&=& 0 = -\pounds_\xi\omega^{31},\nonumber\\
\pounds_\xi\omega^{23}&=& 0 = -\pounds_\xi\omega^{32}.
\end{eqnarray}
As before, we express the perturbation of mass as,
\begin{equation}
m(T,r)=m_{0}+\Delta m (T,r),
\end{equation}
\begin{equation}
\Delta \sqrt{m}= {\Delta m \over 2 \sqrt{m}},
\end{equation}
\begin{equation}
\Delta ({1\over\sqrt{m}})= {-\Delta m \over 2 m \sqrt{m}},
\end{equation}

and obtain the following perturbation of the orthonormal tetrad
and spin-connection respectively,
\begin{eqnarray}
\Delta \vartheta^0&=& 0, \nonumber \\
\Delta \vartheta^1&=&{1\over\sqrt{2m r}}  (\Delta m) d T, \nonumber \\
\Delta \vartheta^2 &=& 0,  \nonumber \\
\Delta \vartheta^3 &=& 0,
\end{eqnarray}
\begin{eqnarray}
\Delta\omega^{01}&=&-\Delta\omega^{10} \nonumber\\
&=& -(\Delta m'- {\Delta m\over r}) \left({1\over \sqrt{2m r}} dr
+{1 \over r} dT \right) \nonumber\\&&+(m'- {m\over r}) {\Delta
m\over 2 m \sqrt{2m r}} dr,\nonumber
\\
\Delta\omega^{02}&=&-{1\over\sqrt{2m r}}  (\Delta m) d\theta
= -\Delta\omega^{20},\nonumber\\
\Delta\omega^{03}&=& -{1\over\sqrt{2m r}}  (\Delta m) \sin\theta d\phi
= -\Delta\omega^{30},\nonumber\\
\Delta\omega^{12}&=& 0 = -\Delta\omega^{21},\nonumber\\
\Delta\omega^{13}&=& 0 =-\Delta\omega^{31},\nonumber\\
\Delta\omega^{23}&=& 0 =-\Delta\omega^{32}.
\end{eqnarray}

Put all the above results into the energy flux expression
$\mathbb{F}(\xi)$,  we obtain the following  nonvanishing
contributions, which includes:
\begin{eqnarray}
&&{1\over 8 \pi}\oint_{\partial\Sigma} i_\xi
(\pounds_\xi\omega^{02}\wedge \Delta
\eta_{02}+\pounds_\xi\omega^{03}\wedge \Delta \eta_{03}) \nonumber\\
&=&-{1\over 2} \Delta m \left( {\tilde {c}_1^2 \dot{m} \over m}
+{\tilde {c}_1 \tilde{c}_2(r) m' \over m} -{\tilde {c}_1
\tilde{c}_2(r) \over  r} \right) ,
\end{eqnarray}

\begin{eqnarray}
&&- {1\over 8 \pi}\oint_{\partial\Sigma} i_\xi (\Delta
\omega^{01}\wedge \pounds_\xi \eta_{01})
\nonumber\\
&=& \tilde{c}_2(r) \left[  - \left(\Delta m'- {\Delta m\over r}
\right) \left({\tilde{c}_2(r)\sqrt{r \over 2m}}
+ \tilde {c}_1 \right) \right] \nonumber\\
&&+ \tilde{c}_2(r) \left[\left(m'- {m\over r} \right)
\left(\tilde{c}_2(r) {\Delta m \over 2m } \sqrt{r\over 2m} \right)
\right] , \label{88}
\end{eqnarray}
and
\begin{eqnarray} \label{89}
&&-{1\over 8 \pi}\oint_{\partial\Sigma} i_\xi (\Delta
\omega^{02}\wedge \pounds_\xi \eta_{02})-{1\over 8
\pi}\oint_{\partial\Sigma} i_\xi (\Delta \omega^{03}\wedge
\pounds_\xi \eta_{03})
\nonumber\\
&=&  { \tilde{c}_2(r) \Delta m \over \sqrt{2mr}}\left(\tilde {c}_1
\sqrt{2m\over r}
+\tilde{c}_2(r)\right) \nonumber\\
&&+{1\over 2} \Delta m \left( {\tilde {c}_1^2 \dot{m} \over m}
+{\tilde {c}_1 \tilde{c}_2(r) m' \over m} -{\tilde {c}_1
\tilde{c}_2(r) \over r} \right) \nonumber\\
&&+ {1\over 2} \Delta m
\tilde{c}_2(r) \tilde{c}_2'(r) \sqrt{r\over 2m} .
\end{eqnarray}
Finally, the total energy flux becomes,
\begin{eqnarray} \label{flux49}
\mathbb{F}(\xi) &=& \Big[ -  \tilde{c}_2(r) \left(\Delta m'- {\Delta
m\over r} \right) \left({\tilde{c}_2(r) \sqrt{r \over 2m}}
+{\tilde {c}_1} \right) \nonumber\\
&&+  {\tilde{c}_2(r) \Delta m \over \sqrt{2mr}} \left( \tilde
{c}_1 \sqrt{2m\over r} +\tilde{c}_2(r) \right) \nonumber\\
&&+ \tilde{c}_2(r)^2 \left(m'- {m\over r} \right) \left({\Delta m
\over 2m } \sqrt{r\over 2m} \right) \nonumber\\
&&+ {1\over 2}
\Delta m \tilde{c}_2(r) \tilde{c}_2'(r) \sqrt{r\over 2m} \Big]_{r_1}^{r_2} .
\end{eqnarray}

Note that an expression on the dynamical horizon was obtained by
Ashtekar and Krishnan \cite{Ashtekar-Krishnan2002} previously,
\begin{eqnarray} \label{Ashtekarflux}
\left(\frac{r_2}{2}- \frac{r_1}{2}\right) &=& \int_{\Delta H}
T_{ab}\tau^{\,a}\xi_{(r)}^b\,d^3V \nonumber\\ && +\frac{1}{16\pi}
\int_{\Delta H} N_r\left\{ |\sigma|^2 + 2|\zeta|^2\right\} \,d^3V
. \nonumber\\
\end{eqnarray}

At the dynamical horizon, $r=2m$, therefore $m'={1/2}, \Delta
m'={1/2}, m/r=1/2$, our flux expression (\ref{flux49}) reduced to
\begin{eqnarray}
\mathbb{F}(\xi) &=& {1\over 2}  \left[\tilde{c}_2(r) (\tilde {c}_1
+\tilde{c}_2(r))(1-{4\Delta m \over r})\right]^{r_2}_{r_1}
\nonumber\\
&&+ {1\over 2} \left[ \Delta m \tilde{c}_2(r) \tilde{c}_2'(r)
\right]^{r_2}_{r_1} .\end{eqnarray} Therefore, on the dynamical
horizon when we choose $\tilde {c}_1=0$ and $ \tilde{c}_2(r) =
\sqrt{r}$ for the displacement vector $\xi$, contribution from the
second term vanishes and our result reduced to an ``area balance
law'',
\begin{eqnarray}
\mathbb{F}(\xi) = ({r_2\over 2} -{r_1 \over 2}),
\end{eqnarray}
which agrees with the Ashtekar-Krishnan energy flux formula
(\ref{Ashtekarflux}), with the shear $|\sigma|$ and the twist
$|\zeta|$ becomes zero in Vaidya spacetime.

Note that similar to the previous section, in this example the
nonvanishing term come from
\begin{eqnarray}
&&\mathbb{F}(\xi) \nonumber\\
&=&-{1\over 16 \pi}\oint_{\partial\Sigma} i_\xi [\Delta
\omega^{01}] \wedge \pounds_\xi (\vartheta^2\wedge \vartheta^3)
\nonumber\\ &&-{1\over 16 \pi}\oint_{\partial\Sigma} (i_\xi
\vartheta^1)[\Delta\omega^{02}\wedge\pounds_\xi\vartheta^3
-\Delta\omega^{03}\wedge\pounds_\xi\vartheta^2]
\nonumber\\
&=&{1\over 16 \pi}\oint_{\partial\Sigma} i_\xi [- \Delta
\omega^{01}+ {\Delta m \over r\sqrt{2mr}}\vartheta^1] \wedge
\pounds_\xi
(\vartheta^2\wedge \vartheta^3), \nonumber\\
\end{eqnarray}
where $\vartheta^2 \wedge \vartheta^3$ is the area element and again hints at
an area dependent first law. For stationary spacetimes
where $\pounds_\xi (\vartheta^2\wedge\vartheta^3)=0$, the energy flux vanishes.

\section{Discussions}

A general expression for quasi-local energy flux expression is derived
from covariant Hamiltonian formulation using
functionally differentiability and symplectic structure invariance, which
is coordinate independent. The energy flux expression is given by the
boundary term. The benefits of using symplectic structure
invariance is to avoid the ambiguity in choosing the correct boundary
terms in the Lagrangian that one begins with. This was a core problem
for many other formalisms for diffeomorphism invariant theories in
the literatures. Another important features of this expression
are the independence of the choice of the canonical variables.
The expression  $\mathbb{F}(\xi)$
depends on the vector field $\xi$ and the choice of the
boundary surfaces which depends on the type of the physics  under
investigation.

For the boundary surface taking to be the null infinity, the
expression leads to the Bondi type energy flux obtained by
Lindquist, Schwartz and Misner \cite{LSM} for Vaidya spacetime,
where the energy flux has the interpretation as the luminosity of
the star as seen by an observer at infinity. If the boundary
surface is taken to be the dynamical horizons, the expression
gives rise to the energy flux obtained by Ashtekar and Krishnan
for Vaidya spacetime.

Note that for Painlev\'{e}-Gullstrand coordinates our expression
fails to give the correct Bondi type energy flux at null infinity
because of the $1/\sqrt{r}$ fall off of the metric. Also for
Schwarzschild coordinates the expression fails on the interesting
dynamical horizon. This only indicates a trivial fact that one
requires to use at least two coordinate patches to cover the whole
spacetime range of interest except for the uninteresting flat
spacetime. However, this is a good news to our energy flux
expression which derived from a coordinate independent covariant
formalism and detects the limits of the applicability of the
coordinate system employed in the calculations.

Another interesting observation is that the expression gives an area dependence which
hints at the first law for general Vaidya spacetime.

\section*{Appendix}
In comparison to the flux expressions in [4],
our flux expression (\ref{15}) is simply the sum of
the Dirichlet and Neumann boundary flux expressions given in [4].
In the special case when $\phi$ is a fixed variable, $\Delta \phi=0$,
our flux expression (\ref{15}) reduced to the
``Dirichlet boundary flux expression'' (equation (30) of [4]),
\begin{equation}
\mathbb{F}_{\rm Dirichlet}(\xi)=
\oint_{\partial\Sigma} i_\xi (\pounds_\xi \phi \wedge \Delta p) .
\end{equation} 
If $p$ is a fixed variable, $\Delta p=0$,
our flux expression (\ref{15}) reduced
to the ``Neumann boundary flux expression''
(equation (31) of [4]),
\begin{equation}
\mathbb{F}_{\rm Neumann}(\xi)=
\oint_{\partial\Sigma} i_\xi (-\Delta \phi \wedge \pounds_\xi p).
\end{equation} 
Similarly, when the spatial projections of the variables $\phi$ and $p$
are fixed, we obtain the ``dynamic boundary flux expression'' (equation (32) of [4]),
\begin{equation}
\mathbb{F}_{\rm dynamic}(\xi)=
\oint_{\partial\Sigma} (\varsigma \pounds_\xi \phi \wedge i_\xi \Delta p
- i_\xi \Delta_\xi \phi \wedge \pounds_\xi p) ,
\end{equation} 
where $\varsigma=(-1)^f$ for $\phi$ being an $f$-form field.
If  the time projections of the variables $\phi$ and $p$
are fixed, we obtain the ``constraint boundary flux expression'' (equation (33) of [4]),
\begin{equation}
\mathbb{F}_{\rm constraint}(\xi)=
\oint_{\partial\Sigma} (i_\xi \pounds_\xi \phi \wedge \Delta p
- \varsigma \Delta_\xi \phi \wedge i_\xi \pounds_\xi p) .
\end{equation} 
Our flux expression (\ref{15}) is the general symplectic invariant expression without any variable
being fixed.

\begin{acknowledgments}
The authors thank J. M. Nester for helpful discussions. RST is
supported
 by NSFC (10375081, 10375087, 10771140) and
Shanghai Education Development Foundation (05SG45).  HLY wants to
thank Department Fisica Fonamental, University of
 Barcelona for hospitality when this work is completed. This work is
 partially supported by National Center for Theoretical Sciences, Taiwan.
\end{acknowledgments}


\end{document}